\documentclass[journal]{IEEEtran}

\usepackage{amsmath,amssymb,amsfonts}
\usepackage{graphicx}
\usepackage{cite}
\usepackage{array}
\usepackage[hidelinks]{hyperref}
\usepackage{textcomp}
\usepackage{xcolor}
\usepackage[caption=false,font=normalsize,labelfont=sf,textfont=sf]{subfig}
\usepackage{url}

\begin{document}

\title{Cross-Modal Characterization of Infant Cry: Validation of a Chest-Surface Accelerometer in Extracting Acoustic Vocal Function Measures}

\author{Winko~W.~An$^{\dagger}$,
        Saketh~Sundar$^{\dagger}$,
        Lisa~Yankowitz,
        Daryush~D.~Mehta,
        and~Carol~L.~Wilkinson%
\thanks{$^{\dagger}$W. W. An and S. Sundar contributed equally to this work.}%
\thanks{Manuscript received TBD date, 2026.}
\thanks{W. W. An and C. L. Wilkinson are with the Division of Developmental Medicine, Boston Children's Hospital, Boston, MA 02115 USA, and also with Harvard Medical School, Boston, MA 02115 USA.}%
\thanks{S. Sundar is with the Division of Developmental Medicine, Boston Children's Hospital, Boston, MA 02115 USA, and also with Harvard University, Cambridge, MA 02138 USA.}%
\thanks{L. Yankowitz is with the Children's Hospital of Philadelphia, Philadelphia, PA 19104 USA.}%
\thanks{D. D. Mehta is with Harvard Medical School, Boston, MA 02115 USA, and also with the Center for Laryngeal Surgery and Voice Rehabilitation, Massachusetts General Hospital, Boston, MA 02114 USA.}%
\thanks{Corresponding author: C. L. Wilkinson.}%
}

\markboth{IEEE Transactions on Audio, Speech, and Language Processing,~Vol.~XX, No.~X, Month~2026}%
{An \MakeLowercase{\textit{et al.}}: Cross-Modal Characterization of Infant Cry}

\maketitle

\begin{abstract}
Infant cry acoustics provide a promising window into early neurodevelopment and may serve as scalable biomarkers for neurodevelopmental disorders. However, conventional microphone-based recordings are highly susceptible to environmental noise and raise privacy concerns in real-world clinical settings. Chest-surface accelerometers may offer a robust alternative by capturing vibrations directly from the larynx. We evaluated the validity of a chest-mounted accelerometer (ACC) for infant cry analysis by comparing acoustic features derived from ACC and simultaneously recorded microphone (MIC) signals during routine vaccination visits. The final sample included 85 infants (41 at 4 months; 44 at 12 months) from a diverse pediatric population. Seven vocal measures were extracted from both modalities, including fundamental frequency (F0), jitter, shimmer, cepstral peak prominence (CPP), and harmonics-to-noise ratio (HNR). Agreement and consistency between modalities were assessed using intraclass correlation coefficients (ICCs). F0 demonstrated excellent agreement between ACC and MIC recordings (ICC $>$ 0.94). Jitter measures also showed good-to-excellent agreement, while CPP demonstrated moderate agreement. Shimmer and HNR showed lower absolute agreement and systematic bias between modalities, reflecting possible differences in signal transmission and noise sensitivity. In summary, chest-surface accelerometers can reliably capture several clinically relevant acoustic features of infant cry, particularly temporal measures of F0 and jitter. This approach offers a noise-robust and privacy-preserving alternative to microphone-based recordings, supporting its potential use in scalable clinical and developmental research applications.
\end{abstract}


\section{Introduction}
\IEEEPARstart{C}{rying} is an important communication mechanism for infants, and infant cries therefore provide a unique window into early brain development. Resulting from the coordinated activity of brainstem, limbic, and cortical motor systems, cry acoustics have been proven to reflect the integrity of neural circuits that will later support speech and social interaction \cite{parkinson2012,newman2007}. Previous research has suggested that acoustic variations in neonatal cries can serve as predictive markers for developmental outcomes. For instance, fundamental frequency (F0) variation has been shown to predict language acquisition at 18 months \cite{shinya2017}. Other acoustic features, such as energy, formants, utterances, voicing, fricatives, and signal quality have also shown modest predictive value for behavioral, language, and cognitive delays \cite{manigault2023}.

Over the past decade, cry characteristics of infant cries have increasingly been investigated as early biomarkers for neurodevelopmental disorders such as Autism Spectrum Disorder (ASD) \cite{esposito2017,english2019}. For example, elevated F0 has been reported across multiple studies in infants later diagnosed with ASD compared to those without later diagnoses \cite{esposito2017,esposito2014,sheinkopf2012,esposito2010}. However, studies focused on infants with and without elevated likelihood for autism due to family history have either reported lower F0 in the newborn period \cite{orlandi2012} or no group differences at 12 months \cite{unwin2017}. Perturbation measures such as jitter and shimmer have been studied less frequently but show a similar pattern: increased jitter and shimmer have been reported in elevated-likelihood infants \cite{santos2013}, and irregularities in these measures have been linked to motor and neurological differences \cite{teixeira2015} as well as distress levels \cite{laguna2023}. Therefore, cry acoustics may offer a scalable and unbiased biomarker of early neurodevelopment.

However, methodological challenges have hindered translation to clinical use. Most prior studies relied on laboratory microphone recordings from small, homogeneous samples of typically $<15$ infants per group \cite{pusil2025}, which limits generalizability to a broader population. One notable exception is Manigault et al., which enrolled 556 infants and recorded cries in neonatal intensive care units. Nevertheless, substantial challenges in data retention remain, with only 65.3\% of recordings deemed usable due to reasons including background noise and insufficient sound quality \cite{manigault2023}. For cry acoustics to serve as a valid clinical biomarker, it must be possible to reliably extract high-quality data in real clinical environments. In practice, recording infant cries in clinical settings is challenging due to substantial environmental noise, including speech from parents or nurses and extraneous sounds such as air conditioning systems or the beeping of hospital equipment \cite{ji2021}. These factors can lead to unreliable estimates of acoustic parameters and reduced reproducibility across recordings. Additionally, audio recordings often contain identifiable or sensitive information about participants, particularly when infant cries occur in the presence of caregivers or clinical staff \cite{wiepert2024}. This raises privacy concerns that further limit the feasibility of sharing large audio datasets publicly, thereby constraining the development and application of large computational models for biomarker discovery.

To address these issues, recent advances in sensor technology enable direct measurement of voice-related vibrations using lightweight accelerometers placed on the skin at the suprasternal notch. These neck- or chest-surface accelerometers capture subglottal vibrations generated at the larynx and produce a cleaner signal that is relatively robust to environmental noise \cite{cortes2022,mehta2012,mehta2016}. The approach has been well validated in adult voice research across a range of tasks, with strong correlations between features derived from accelerometer signals and those extracted from microphone recordings \cite{mehta2016,cantorcutiva2025}. In addition to its consistency with, and technical advantages over, traditional audio recordings, this method is affordable, portable, and potentially straightforward to implement in infants.

In this study, we investigated the use of chest-surface accelerometers for estimating typical acoustic features from infant cry, similar to a prior neck-surface vibration study with adult speakers \cite{mehta2016}. Specifically, we collected concurrent microphone (MIC) and accelerometer (ACC) recordings during vaccine injection--induced crying episodes in infants aged 4 and 12 months. We annotated cry segments that were free of salient environmental noise and extracted vocal function measures, including fundamental frequency (F0), jitter, shimmer, and related vocal parameters. We then compared vocal measures derived from MIC and ACC signals and assessed their agreement and consistency using intraclass correlation coefficients (ICCs).

\section{Methods}

\subsection{Data Collection}
The Boston Children’s Hospital Institutional Review Board reviewed and approved this study (IRB-P00038922). During routine 4- and 12-month vaccination visits, we obtained high-quality microphone and synchronized recordings from accelerometers \cite{mehta2012}. This method produces recordings of standardized, pain-triggered cries and fits practically into clinical workflows. The final sample included recordings from 41 4-month-old (24 female, 17 male) and 44 12-month-old (22 female, 22 male) infants, randomly selected from a larger pool of participants. Each session was conducted in an examination room at the Boston Children's Hospital Primary Care Clinic, which serves a racially, ethnically, and socioeconomically diverse patient population (approximately 85\% Black or Hispanic, 65\% publicly insured). The natural clinical environment allowed for testing sensor validity amid common acoustic interference in clinical settings, such as medical equipment alerts and multiple speakers in the environment.

\begin{figure*}[!t]
\centering
\subfloat[]{\includegraphics[width=0.45\textwidth]{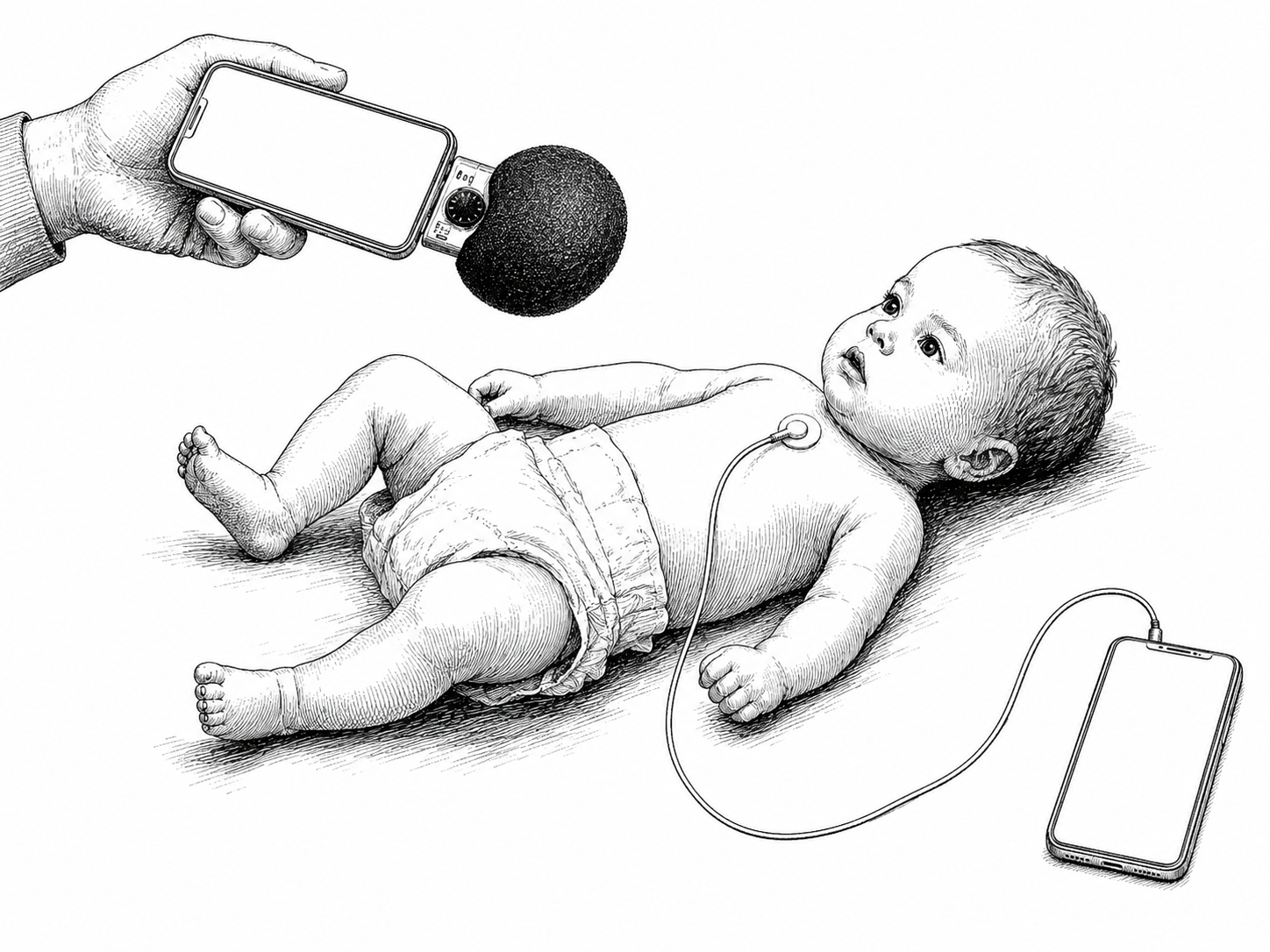}}
\hfil
\subfloat[]{\includegraphics[width=0.35\textwidth]{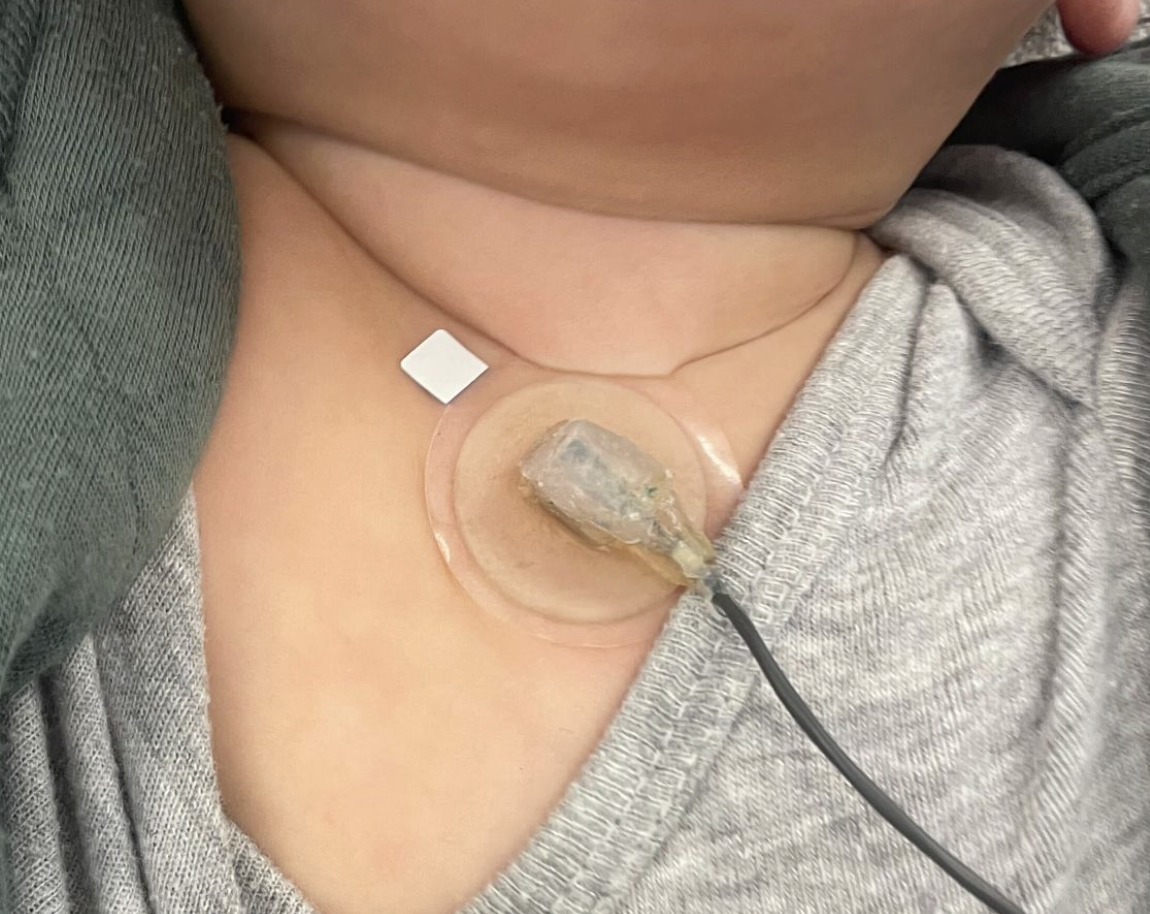}}
\caption{(a) Illustration of sensor attachment and equipment used in the experiment. (b) Close-up photo of the accelerometer placement on an infant.}
\label{fig:setup}
\end{figure*}

Fig.~\ref{fig:setup} illustrates the recording setup. To record voicing-related vibrations, the chest-surface accelerometer (ACC; Knowles BU-27135) was secured slightly above the suprasternal notch using hypoallergenic adhesive tape and continuously recorded at 11.025~kHz to an Android smartphone \cite{mehta2012}. An iPhone with a Zoom iQ7 stereo microphone (MIC; 44.1~kHz sampling rate) was used to record audio simultaneously from roughly 0.2~meters away, though the distance was not standardized. Recordings started shortly before the first injection, with timestamps logged for each injection. Recordings lasted for 90~seconds or until the crying stopped. For each recording, sensor positioning, any unusual background noise such as music or videos played by the caregiver, and observations of infant positioning were documented.

\subsection{Data Labeling}
\begin{figure*}[!t]
\centering
\includegraphics[width=\textwidth,keepaspectratio]{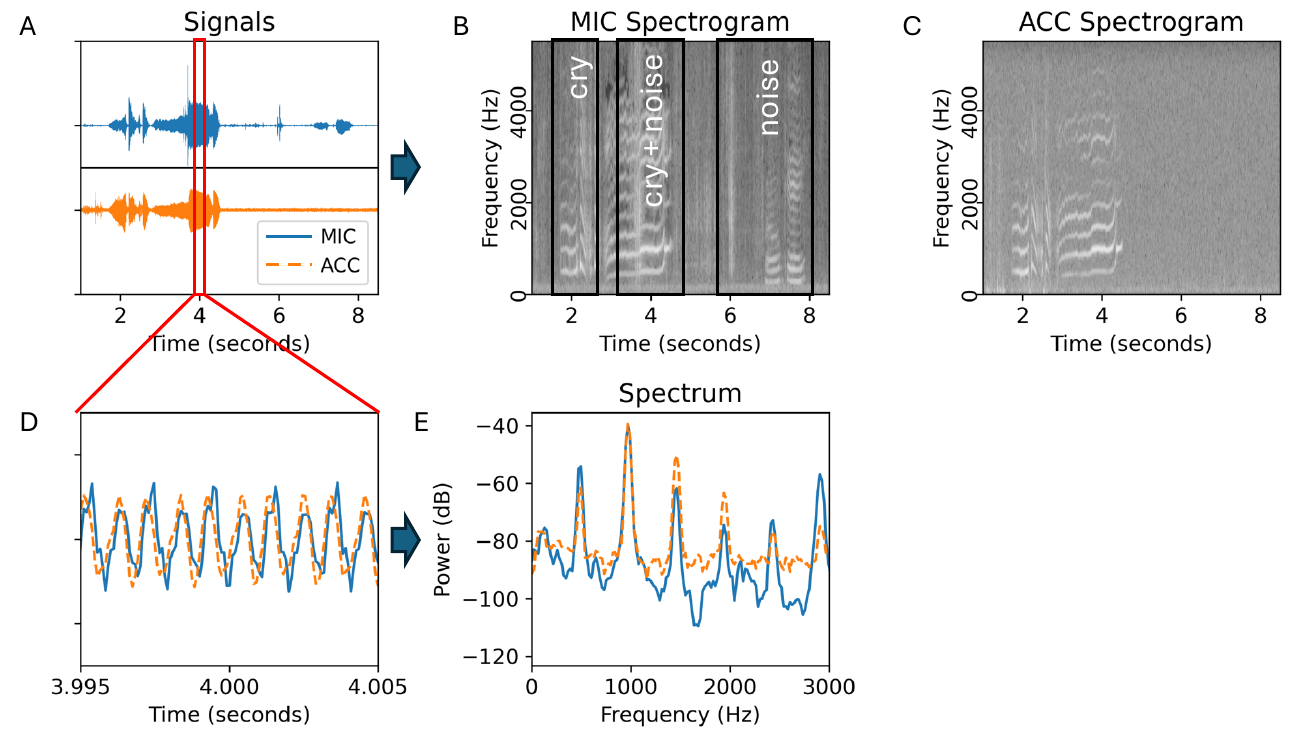}
\caption{(a) Raw microphone (MIC; blue) and accelerometer (ACC; orange) waveforms from a single recording containing cry and non-cry segments. The red box indicates the region shown in panel (d). (b) Spectrogram of the MIC signal, with labeled segments corresponding to cry (cry-only), cry mixed with noise (cry+noise), and noise only (non-cry). (c) Spectrogram of the ACC signal over the same time period. (d) Zoomed-in view ($\sim$10~ms) of MIC and ACC waveforms, showing individual glottal cycles. (e) Power spectral density of the MIC and ACC signals.}
\label{fig:annot}
\end{figure*}

After downsampling the MIC signal to 11.025~kHz and synchronizing the MIC and ACC signals using a cross-correlation method \cite{mehta2016}, every recording was examined and annotated manually using Praat \cite{praat}. Based on the presence of infant vocalization and characteristics of background noise, the recordings were annotated with three labels: (1) cry-only: segments with distinct infant cries that did not have background noise; (2) cry+noise: segments with infant cries that overlapped with speech of others/background noise; and (3) non-cry: segments that did not have distinct cry audio. Additionally, to ensure a high signal-to-noise ratio, any segment with a Root-Mean-Square (RMS) amplitude below 0.01 in the MIC signal was excluded from analysis. Fig.~\ref{fig:annot} shows example annotated segments for cry-only and non-cry labels.

\subsection{Data Analysis}
All signal processing and feature extraction were performed in Python (v3.10) using NumPy, SciPy, pandas, and Parselmouth, a Python interface to Praat software \cite{parselmouth}. Each manually labeled cry-only segment was further subdivided into contiguous, non-overlapping 50~ms windows to ensure the presence of multiple glottal cycles for reliable estimation.

We selected seven standard vocal measures to characterize the glottal source characteristics and periodicity of the cry signal independently from the ACC and MIC signal from the same cry. The measures were chosen based on their prior use in adult ACC--MIC comparisons and/or their use in past studies of cry acoustics. Although the traditional measures of jitter and shimmer are no longer part of the recommended battery of clinical voice measures \cite{patel2018}, their application in the current study serves a different purpose for sensor validation. Jitter and shimmer are fine timescale measures of periodicity and amplitude, respectively, that are meant to characterize and validate the ACC signal relative to the MIC signal.

\subsubsection{Fundamental Frequency (F0)}
F0 reflects the rate of vocal fold vibration. For each recording, F0 was estimated from the entire recording using Praat's cross-correlation method (\texttt{to\_pitch\_cc}) with a time step of 1~ms, a pitch floor of 200~Hz, and a ceiling of 1500~Hz. These parameters were specifically tuned to accommodate the high-pitched nature of infant vocalizations and the high intensity of pain-triggered cries. This pitch range was selected to capture the frequency values where both MIC and ACC signals showed valid voicing.

\subsubsection{Jitter (F0 Perturbation)}
Jitter indexes cycle-to-cycle variability in F0. To ensure robustness, periods were filtered using an Interquartile Range (IQR) method to remove outliers and smoothed with a 3-point uniform filter before calculation. Two measures were computed. The Jitter Coefficient of Variation ($J_{CV}$), defined in \eqref{eq:jcv}, was calculated as the coefficient of variation of the pitch contour, with $p_i$ representing the period of the $i^{\text{th}}$ glottal cycle and $\bar{p}$ representing the average period duration across $N=3$ cycles. The Jitter Local ($J_{\text{local}}$) measure, defined in \eqref{eq:jlocal}, was computed as the average absolute difference between consecutive periods divided by the mean period.

\begin{equation}
\label{eq:jcv}
J_{CV} = \frac{1}{\bar{p}} \sqrt{\frac{1}{N-1}\sum_{i=1}^{N-1}(p_i - \bar{p})^2}
\end{equation}

\begin{equation}
\label{eq:jlocal}
J_{\text{local}} = \frac{1}{\bar{p}(N-2)} \sum_{i=1}^{N-2} |p_{i+1} - p_i|
\end{equation}

\subsubsection{Shimmer (Amplitude Perturbation)}
Shimmer measures the cycle-to-cycle variability in the peak-to-peak amplitude of the glottal pulses. Amplitudes were extracted using a PointProcess (periodic cross-correlation) to identify individual glottal cycles. The Shimmer CV ($S_{CV}$), defined in \eqref{eq:scv}, is the coefficient of variation of the peak amplitudes, with $a_i$ representing the amplitude of the $i^{\text{th}}$ glottal cycle and $\bar{a}$ representing the average amplitude across $N=3$ cycles. The Shimmer Local ($S_{\text{local}}$), defined in \eqref{eq:slocal}, is the mean absolute difference between consecutive amplitudes divided by the mean amplitude.

\begin{equation}
\label{eq:scv}
S_{CV} = \frac{1}{\bar{a}} \sqrt{\frac{1}{N}\sum_{i=1}^{N}(a_i - \bar{a})^2}
\end{equation}

\begin{equation}
\label{eq:slocal}
S_{\text{local}} = \frac{1}{\bar{a}(N-1)} \sum_{i=1}^{N-1} |a_{i+1} - a_i|
\end{equation}

\subsubsection{Cepstral Peak Prominence (CPP)}
CPP reflects the degree of harmonic organization in the voice and is a widely used measure of voice quality and dysphonia severity \cite{patel2018,murton2020}. To calculate the CPP, each 50~ms cry segment was transformed into a log-power spectrum and then into a cepstrum using a 40.96~ms Hamming window. A linear regression baseline was fitted across the cepstral power spectrum after applying a 0.67~ms lifter to isolate vocal periodicity from vocal tract information. The peak search was limited to quefrencies between 0.67~ms and 5~ms, corresponding to fundamental frequencies of 1500~Hz and 200~Hz, respectively. The final CPP value was defined as the distance in decibels (dB) between the highest cepstral peak and the corresponding point on the baseline.

\subsubsection{Harmonics-to-Noise Ratio (HNR)}
HNR quantifies the ratio of periodic to aperiodic energy. This was calculated using Praat's time-domain harmonicity analysis via cross-correlation periodicity detection.

\subsection{Statistical Analysis}
To evaluate the agreement between vocal function features derived from the ACC and the MIC signals, we performed a comparative analysis across the 4-month and 12-month age cohorts. We first excluded segments containing outliers, defined as any measure exceeding three standard deviations from the grand average. We randomly selected 20 segments from each infant's recording and averaged values from each vocal function metric within each participant for the analyses below.

We calculated Intraclass Correlation Coefficients (ICCs) using the \texttt{pingouin} statistical library \cite{shrout1979} to quantify the degree of agreement and consistency between the two sensors (ACC and MIC). Values were averaged across all segments within each participant prior to the ICC calculation. Since only one sensor would be used in future applications, we reported ICC(A,1) to assess the absolute agreement between sensors and ICC(C,1) to evaluate the consistency between sensors. Notably, ICC(A,1) is synonymous with the traditional convention of ICC(3,1) \textit{absolute agreement terminology} \cite{shrout1979} for two-way mixed effects \cite{mcgraw1996} with single rater; ICC(C,1) is synonymous with the traditional convention of ICC(3,1) \textit{consistency terminology} for two-way mixed effects with single rater. We followed the guidelines proposed by Koo and Li \cite{kooli2016}, classifying ICC values $<0.50$ as poor, 0.50--0.75 as moderate, 0.75--0.90 as good, and $>0.90$ as excellent agreement. To investigate systematic bias in measures with poor-to-moderate absolute agreement (ICC(A,1)) < 0.75, we conducted subject-level comparisons between ACC and MIC for each metric using paired t-tests.

\section{Results}

\subsection{Fundamental Frequency}
\begin{figure*}[!t]
\centering
\includegraphics[width=\textwidth,keepaspectratio]{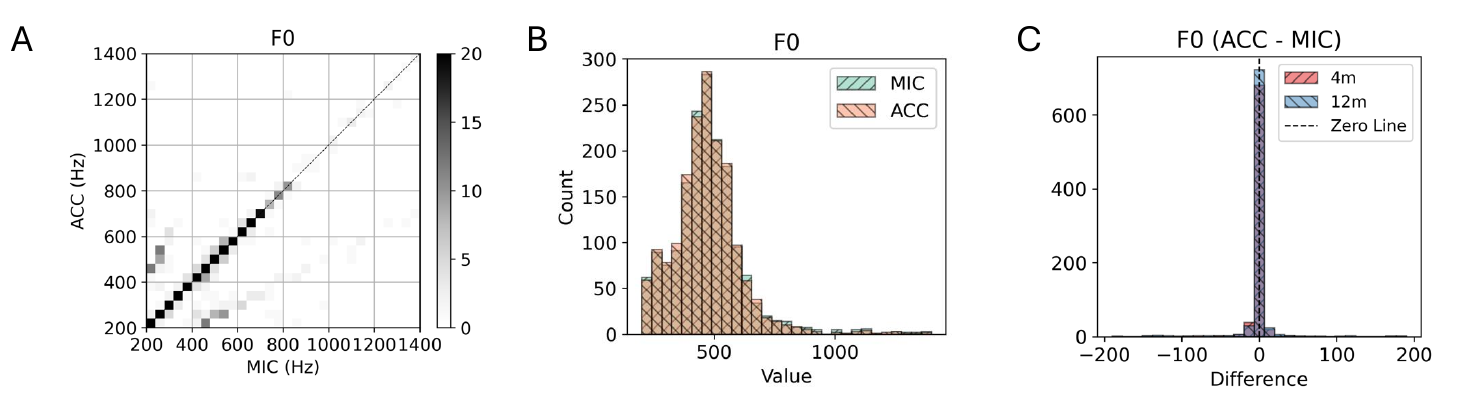}
\caption{Graphical agreement in F0 between microphone (MIC) and accelerometer (ACC) signals from each 50-ms segment. (a) A 2-D histogram illustrating the agreement. (b) 1-D histograms showing distributions of F0 in the MIC and ACC signals. (c) Distribution of within-segment differences between MIC and ACC measures of F0, separated by age.}
\label{fig:f0}
\end{figure*}

F0 extracted from MIC and ACC signals demonstrated excellent agreement; the vast majority of tested segments showed nearly identical F0 values (Fig.~\ref{fig:f0}), with only a small proportion of data points exhibiting discrepancies attributable to pitch halving ($\sim$2.7\%) or doubling ($\sim$3.5\%) (Fig.~\ref{fig:f0}(a)). Consistent with this observation, ICC analyses indicated excellent absolute agreement and consistency between the two signals (Table~\ref{tab:icc}), which is true in both age groups.

\begin{table*}[!t]
\caption{Intraclass Correlation Coefficients Between MIC and ACC Signals}
\label{tab:icc}
\centering
\renewcommand{\arraystretch}{1.8}
\begin{tabular}{lcccccc}
\hline\hline
& \multicolumn{2}{c}{Overall} & \multicolumn{2}{c}{4 months} & \multicolumn{2}{c}{12 months}\\
\cline{2-3}\cline{4-5}\cline{6-7}
Measure & ICC(A,1) & ICC(C,1) & ICC(A,1) & ICC(C,1) & ICC(A,1) & ICC(C,1)\\
\hline
F0 (Hz)                 & \textbf{0.947} & \textbf{0.950} & \textbf{0.942} & \textbf{0.950} & \textbf{0.954} & \textbf{0.954}\\
$J_{CV}$ (\%)           & \textbf{0.949} & \textbf{0.958} & \textbf{0.959} & \textbf{0.965} & \textbf{0.919} & \textbf{0.935}\\
$J_{\text{local}}$ (\%) & \textbf{0.873} & \textbf{0.872} & \textbf{0.903} & \textbf{0.901} & \textbf{0.819} & \textbf{0.817}\\
$S_{CV}$ (\%)           & 0.187 & 0.647 & 0.208 & 0.700 & 0.154 & 0.577\\
$S_{\text{local}}$ (\%) & 0.322 & 0.601 & 0.320 & 0.594 & 0.309 & 0.589\\
CPP (dB)                & 0.583 & 0.586 & 0.598 & 0.593 & 0.573 & 0.584\\
HNR (dB)                & 0.411 & 0.610 & 0.437 & 0.638 & 0.378 & 0.572\\
\hline\hline
\multicolumn{7}{l}{\footnotesize Bold indicates good-to-excellent agreement or consistency (ICC $> 0.75$).}
\end{tabular}
\end{table*}

\subsection{Jitter}
\begin{figure*}[!t]
\centering
\includegraphics[width=\textwidth,keepaspectratio]{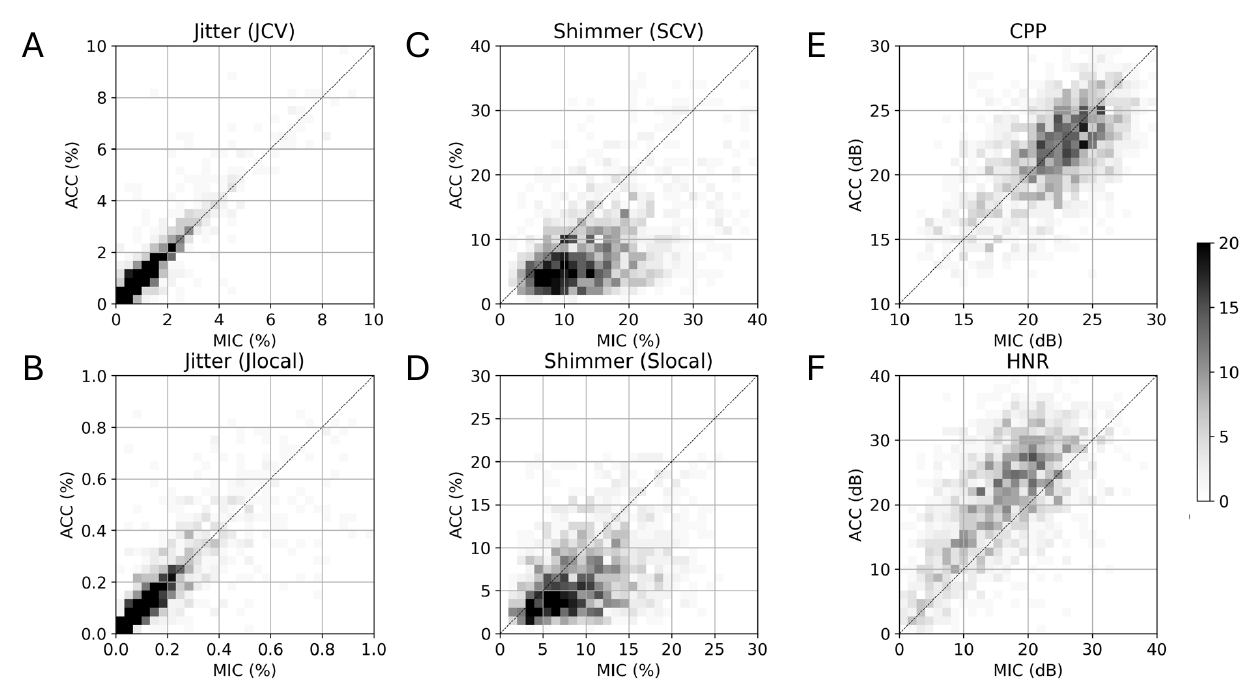}
\caption{Two-dimensional histograms visualizing agreement in vocal function measures between microphone (MIC) and accelerometer (ACC) signals from each 50-ms segment. (a) Jitter Coefficient of Variation ($J_{CV}$). (b) Jitter Local ($J_{\text{local}}$). (c) Shimmer Coefficient of Variation ($S_{CV}$). (d) Shimmer Local ($S_{\text{local}}$). (e) Cepstral Peak Prominence (CPP). (f) Harmonics-to-Noise Ratio (HNR).}
\label{fig:histo2d}
\end{figure*}

\begin{figure*}[!t]
\centering
\includegraphics[width=\textwidth,keepaspectratio]{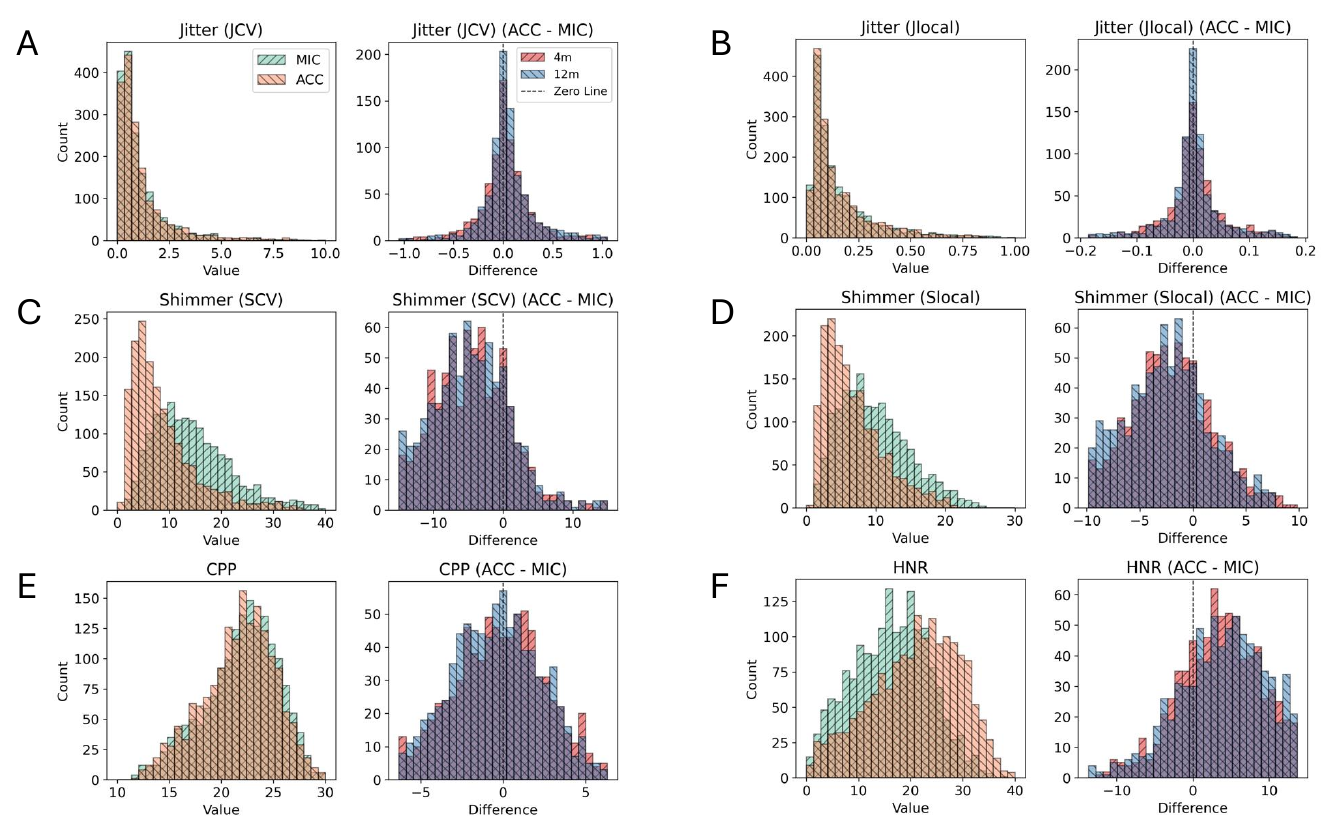}
\caption{Histograms showing distribution of vocal function values extracted in microphone (MIC) and accelerometer (ACC) signals from each 50-ms segment and their differences, separated by age group. (a) Jitter Coefficient of Variation ($J_{CV}$). (b) Jitter Local ($J_{\text{local}}$). (c) Shimmer Coefficient of Variation ($S_{CV}$). (d) Shimmer Local ($S_{\text{local}}$). (e) Cepstral Peak Prominence (CPP). (f) Harmonics-to-Noise Ratio (HNR).}
\label{fig:histo1d}
\end{figure*}

Overall absolute agreement and consistency in jitter between sensor types was good-to-excellent (Figs.~\ref{fig:histo2d} and~\ref{fig:histo1d}; Table~\ref{tab:icc}). Between the two jitter measures, agreement was generally higher for $J_{CV}$ than for $J_{\text{local}}$ at both age points (Table~\ref{tab:icc}), consistent with findings in a prior study in adults \cite{mehta2016}.

\subsection{Shimmer}
Both shimmer metrics demonstrated generally moderate consistency between sensor modalities, but poor absolute agreement (Figs.~\ref{fig:histo2d} and~\ref{fig:histo1d}; Table~\ref{tab:icc}). Notably, shimmer values derived from the ACC signal were consistently lower than those obtained from the MIC signal (Figs.~\ref{fig:histo2d} and~\ref{fig:histo1d}; Table~\ref{tab:bias}). A similar trend was reported in a prior study in adults \cite{mehta2012,mehta2016}, although the magnitude of the difference was smaller than observed here ($S_{CV}$ bias $\sim$2 percentage points [pp]; $S_{\text{local}}$ bias $\sim$1~pp).

\subsection{Cepstral Peak Prominence}
Both absolute agreement and consistency were moderate for CPP (Table~\ref{tab:icc}). However, the bias in CPP seemed to vary by age. At 4 months, the difference between modalities was insignificant (Table~\ref{tab:bias}). By 12 months, however, the CPP values extracted from the ACC signal were marginally lower than those obtained from the MIC signal (Table~\ref{tab:bias}). The negative bias was also found in previous work in adults in the /a/ vowel condition, which is a similar vocal tract configuration to the infant cry (compared with /i/ and /u/).

\subsection{Harmonics-to-Noise Ratio}
HNR demonstrated moderate consistency and poor absolute agreement between modalities (Table~\ref{tab:icc}). HNR values derived from the ACC signal were substantially higher (4--5~dB) than those obtained from the MIC signal (Fig.~\ref{fig:histo1d}; Table~\ref{tab:bias}), in the same direction as findings from the previous study in adult production of /a/ vowels of a 7~dB bias for the ACC signal \cite{mehta2016}. The infant cry positive ACC bias for HNR was observed in both the 4-month (4.38~dB) and 12-month (5.11~dB) age groups.

\begin{table}[!t]
\caption{Average Bias from the ACC Signal with MIC Measure as Reference}
\label{tab:bias}
\centering
\renewcommand{\arraystretch}{1.8}
\begin{tabular}{lcccc}
\hline
& \multicolumn{2}{c}{4m} & \multicolumn{2}{c}{12m}\\
\cline{2-3}\cline{4-5}
Measure & Bias & $p$ & Bias & $p$\\
\hline
$S_{CV}$ (pp)           & \textbf{$-$5.983} & \textbf{7.87 $\times$ 10$^{-21}$} & \textbf{$-$6.603} & \textbf{1.32 $\times$ 10$^{-20}$}\\
$S_{\text{local}}$ (pp) & \textbf{$-$2.803} & \textbf{1.27 $\times$ 10$^{-11}$} & \textbf{$-$3.250} & \textbf{1.17 $\times$ 10$^{-12}$}\\
CPP (dB)                & $-$0.079 & 7.51 $\times$ 10$^{-1}$ & $-$0.430 & 9.00 $\times$ 10$^{-2}$\\
HNR (dB)                & \textbf{4.381} & \textbf{7.20 $\times$ 10$^{-9}$} & \textbf{5.110} & \textbf{4.03 $\times$ 10$^{-9}$}\\
\hline
\multicolumn{5}{p{0.95\columnwidth}}{\footnotesize ACC minus MIC. Negative bias indicates the ACC signal is lower than the MIC signal. Bold indicates statistically significant difference ($p < 0.05$).}
\end{tabular}
\end{table}

\section{Discussion}
We evaluated the validity of using a chest-surface vibration sensor to record the vocal characteristics of infant cries and extract clinically relevant acoustic features. By directly comparing measures derived from the vibration sensor signal to those derived from simultaneously recorded audio signals, we demonstrated varied consistency and agreement between sensor modalities across features. Timing-based metrics of F0 and jitter measures showed excellent absolute agreement; cepstral peak prominence (CPP) showed moderate absolute agreement and consistency; harmonics-to-noise ratio (HNR) demonstrated moderate absolute agreement with a negative bias for the ACC modality. As expected, amplitude-based metrics of shimmer exhibited the least agreement between sensor modalities, which was demonstrated in the previous study with adult speakers \cite{mehta2012,mehta2016}. Together, these findings validate the use of a chest-surface accelerometer sensor as a reliable alternative to traditional microphone-based recordings for infant cry analysis.

Among all measures examined, F0 demonstrated the strongest agreement between sensor modalities. This finding is particularly important in the context of neurodevelopmental disorder research. Atypical cry pitch characteristics, including altered mean F0 and F0 variability, have been repeatedly associated with developmental risk for autism spectrum disorder \cite{esposito2017,esposito2014,sheinkopf2012,esposito2010} and other neurodevelopmental conditions \cite{lagasse2005,lester2002}. Reliable capture of F0 using a chest-surface sensor thus provides a robust foundation for early biomarker development. Notably, the previous adult study \cite{mehta2016} also reported extremely high correlations for F0 between MIC and ACC signals, reinforcing that timing-related vocal measures translate well across sensor modalities regardless of age.

Beyond F0, other acoustic features demonstrated varied agreement across modalities, and systematic biases were observed in some. Specifically, shimmer values derived from the ACC signal were consistently lower than those obtained from the MIC signal, whereas HNR tended to be higher in the ACC signal, both suggesting greater signal stability in ACC recordings compared with MIC. Nevertheless, the overall moderate consistency agreement supports the use of ACC-derived measures for comparing acoustic features across populations, such as autism versus non-autism. These modality-dependent differences likely reflect the distinct physical pathways of signal transmission \cite{mehta2016}. Chest-surface sensors capture tissue-conducted vibrations and are inherently less influenced by radiated acoustic noise and environmental interference \cite{zanartu2009}, which may enhance measurement stability, particularly in naturalistic infant recordings. The low ICCs obtained for the shimmer metrics are potentially a consequence of the ACC sensor being less sensitive than the MIC sensor to turbulent acoustic noise sources at the level of the larynx. Importantly, acoustic perturbation and spectral measures have been far less extensively investigated in infant cry and neurodevelopmental disorder research compared with F0. Existing literature on the association between developmental outcomes and perturbation or spectral measures remains limited \cite{santos2013,teixeira2015}. The accelerometer-based method proposed in this study may help address this research gap by providing a viable, scalable, and potentially advantageous approach for quantifying these features in infant populations at scale.

The chest-mounted vibration sensor offers several advantages over conventional microphone-based recording, particularly in infant and clinical contexts. First, privacy protection is substantially improved. Unlike audio recordings, which may capture identifiable speech (e.g., caregiver voices or names), the vibration signal does not contain intelligible linguistic content. This feature is especially valuable for large-scale or home-based monitoring studies involving infants. Second, recording geometry is more controlled. In clinical settings, the distance between microphone and infant can vary substantially due to movement, positioning, or caregiver handling. Such variability can affect amplitude- and spectral-based measures. In contrast, the sensor-to-skin distance remains fixed, providing consistent signal capture across sessions and participants (similar to the consistent placement of a stethoscope). Third, the sensor is less susceptible to environmental noise. As demonstrated in prior work with adult speakers \cite{mehta2016}, chest-surface acceleration is inherently resistant to ambient acoustic interference. This property is particularly advantageous in the study of infant cry, as caregivers often vocalize to comfort their crying infant, and for naturalistic recordings in homes or busy clinical environments. In addition to enabling more reliable extraction of spectral and perturbation measures, the use of accelerometers also allows for more robust characterization of temporal features of infant cry, such as cry interval, cry rate, and pause ratio, which can be substantially distorted by environmental noise in microphone recordings.

The use of accelerometers for infant cry analysis is also subject to several limitations. First, accelerometers do not capture all aspects of crying. Rather than recording the acoustic signal directly, they measure mechanical vibrations at the body surface and may therefore miss features shaped by the motor control and physiology of the vocal tract, including the oral pharynx, tongue, palate, and nasal structures. Future work could partially mitigate this limitation by placing sensors higher on the neck or on the cheek, which may better capture vocal-tract-related changes while retaining robustness to environmental noise. Second, this approach requires attaching a sensor to the infant's chest. Compared to conventional microphone-based recording, it necessitates specialized hardware, proper sensor placement, and dedicated smartphone software, which together increase the practical barriers to implementation. However, the trade-off of yielding higher quality vocal function measurement may outweigh the limitations depending on the application area.

\section{Conclusion}
This study demonstrates that a chest-surface vibration sensor can reliably capture infant cry signals and produce vocal function measures that closely correspond to those derived from conventional microphone recordings. Fundamental frequency related measures showed particularly strong agreement across modalities, whereas amplitude perturbation and spectral measures demonstrated moderate consistency agreement with a general systematic bias. Our findings support the feasibility of using chest-placed accelerometers for scalable cry analysis in infants, which may facilitate the development of large-scale cry datasets and improve the reliability of acoustic biomarkers for early identification of neurodevelopmental risk.

\section*{Acknowledgments}
The authors sincerely thank the infants and families who participated in this study for their time and contribution to this research. The authors also thank the students and research assistants involved in participant recruitment and data collection, and Margaret Shilling who helped with cry annotation. The authors acknowledge the use of OpenAI's DALL\textperiodcentered E image generation model to generate the illustrative graphic in Fig.~1(a), which was edited by the authors for schematic scientific presentation. All technical content, analysis, and conclusions are the authors' own. This study was supported by the Eagles Autism Foundation, NINDS (R01NS120986-01A1), NIDCD (P50 DC015446), and the Boston Children's Hospital Division of Developmental Medicine Seed Grant. W.W.A. received support from the Rosamund Stone Zander Hansjoerg Wyss Translational Neuroscience Center at Boston Children's Hospital. The article’s contents are solely the responsibility of the authors and do not necessarily represent the official views of the National Institutes of Health.

\end{document}